\documentclass[aps,prb,twocolumn,superscriptaddress,longbibliography]{revtex4-1}
\usepackage{amsmath, amssymb}
\usepackage{graphicx}
\usepackage{color}
\usepackage{dcolumn} 

\usepackage[english]{babel}
\usepackage{letltxmacro}

\LetLtxMacro{\ORIGselectlanguage}{\selectlanguage}
\makeatletter
\DeclareRobustCommand{\selectlanguage}[1]{%
  \@ifundefined{alias@\string#1}
    {\ORIGselectlanguage{#1}}
    {\begingroup\edef\x{\endgroup
       \noexpand\ORIGselectlanguage{\@nameuse{alias@#1}}}\x}%
}
\newcommand{\definelanguagealias}[2]{%
  \@namedef{alias@#1}{#2}%
}
\makeatother

\definelanguagealias{en}{english}

\begin{document}
\title{Fractional Hofstadter States in Graphene on Hexagonal Boron Nitride}
\author{Ashley M. DaSilva}
\affiliation{Department of Physics, The University of Texas at Austin, Austin, Texas 78712-1192, USA}
\author{Jeil Jung}
\affiliation{Department of Physics, University of Seoul, Seoul 02504, Korea}
\author{Allan H. MacDonald}
\affiliation{Department of Physics, The University of Texas at Austin, Austin, Texas 78712-1192, USA}

\begin{abstract}
In fractionally filled Landau levels there is only a small energy difference between broken translational symmetry electron-crystal states and 
exotic correlated quantum fluid states.  We show that the spatially periodic substrate interaction 
associated with the long period moir{\' e} patterns present in graphene on nearly
aligned hexagonal boron nitride (hBN) tilts this close competition in favor of the former,
explaining surprising recent experimental findings.  
\end{abstract}
\maketitle
\noindent

\noindent
{\em Introduction}--- The quantum Hall effect is a transport anomaly
in two-dimensional electron systems that is associated with the appearance of charge gaps, {\it i.e.} chemical potential jumps, at 
densities that depend on magnetic field.\cite{thouless_quantized_1982,akkermans_mesoscopic_1995}  
Experimentally the quantum Hall effect is manifested by longitudinal resistances that vanish in the low temperature 
limit, and by Hall conductivities that accurately approach multiples
 of the quantum unit of conductance: $\sigma_{H} \to \sigma e^2/h$ where $\sigma$ is a dimensionless rational number.  
The quantum Hall effect is enriched when electrons respond to interactions,\cite{prange_quantum_1987,jain_composite_2007} 
 by forming fractional quantum Hall states,\cite{tsui_two-dimensional_1982} and when they respond to an external periodic potential,
by forming Hofstadter butterfly states.\cite{wannier_dynamics_1962, hofstadter_energy_1976}   
Recent progress\cite{tang_precisely_2013,woods_commensurate-incommensurate_2014,kim_van_2016} in controlling the alignment between graphene sheets and a hexagonal boron nitride (hBN) substrate has made it possible for the first time to study systems in which interactions and periodic potentials simultaneously play an essential role.  
The small difference between the lattice constants of graphene and hBN leads to a moir{\'e} pattern,\cite{xue_scanning_2011,decker_local_2011,yankowitz_emergence_2012} and to an associated perfectly periodic substrate-interaction correction to the isolated graphene sheet Hamiltonian. 
In this Letter, we examine the close competition between competing correlated fluid and broken-translational symmetry electron-crystal states in fractionally filled Landau levels. We show that the periodic Hamiltonian induced by the moir\'e pattern favors the latter of these two states.

Many aspects of the quantum Hall effect can be understood using simple arguments\cite{macdonald_landau-level_1983} that we briefly summarize below.
The densities at which gaps appear depend on both the density scale associated with the external magnetic field $n_{\Phi} = B/\Phi_0$, and the density scale associated with the external periodic potential, $n_{\rm P}=1/A_0$. 
Here $B$ is the perpendicular magnetic field strength, $\Phi_0$ is the electron magnetic flux quantum, 
and $A_0$ is the unit cell area defined by the translational symmetry of the periodic potential.  
It follows from scaling considerations that the density at which any gap that persists over a finite range of system parameters appears must satisfy $\nu_{\rm P} \equiv n/n_{\rm P} = f(\Phi/\Phi_0)$. 
Here $\nu_{\rm P}$ is the number of electrons per unit cell, $\Phi = A_0 B$ is the flux per unit cell, and $\Phi/\Phi_0=n_{\Phi}/n_{\rm P}$ and $f$ is a function of $\Phi/\Phi_0$.  
(The Landau level filling factor $\nu \equiv n/n_{\Phi} = \nu_{\rm P} \Phi_0/\Phi$.)  
In the absence of interactions, simple single-particle state counting requires that both 
\begin{equation} 
\label{eq:s}
s \equiv \frac{\partial (A n)}{\partial (A/A_0)} = - A_0^2  \frac{ \partial n}{\partial A_0} = \nu_{\rm P} - \frac{\Phi}{\Phi_0} \, f'(\Phi/\Phi_0),
\end{equation}  
and 
\begin{equation} 
\label{eq:sigma}
\sigma \equiv \frac{\partial (A n)}{\partial (A n_{\rm P})} = \Phi_0 \frac{\partial{n}}{\partial B} = f'(\Phi/\Phi_0)
\end{equation} 
be integers, where $A$ is the system area and $f'$ is the derivative of $f$. In Eqs.~\ref{eq:s} and ~\ref{eq:sigma}, $s$ and $\sigma$ are the number of electrons added below the gap for each unit cell area added to the system and for each added quantum of magnetic flux, respectively.  
The gaps of ordinary Bloch band insulators are characterized by integer values of $s$ with $\sigma=0$, whereas the gaps between Landau levels in the integer quantum
Hall effect are characterized by integer values of $\sigma$ with $s=0$. 
The property that the dimensionless Hall conductance $\sigma$ is a readily measured observable\cite{akkermans_mesoscopic_1995} 
is a truly remarkable aspect of quantum Hall physics. 

Comparing Eqs.~\ref{eq:s} and ~\ref{eq:sigma} we see that inside a gap the trajectory of $\nu_{\rm P}$ as $\Phi/\Phi_0$ is varied must be a straight line with a quantized $B=0$ intercept and a quantized slope:
\begin{equation} 
\nu_{\rm P} = s + \sigma \frac{\Phi}{\Phi_0}.
\end{equation}
When interactions are included, low energy states containing no quasiparticle excitations may repeat only after every $m$ magnetic flux quanta and only after every 
$N$ unit cells so that $\sigma = \Sigma/m$ and $s = S/N$ where $\Sigma,m,S,$ and $N$ are all integers.  
(The allowed values of $s$ and $\sigma$ are therefore restricted not to integers but to rational numbers.)  

The integer quartet $(\Sigma,m,S,N)$ that characterizes a gap at a given value of $\nu_{\rm P}$ and $\Phi/\Phi_0$ is not uniquely specified\footnote{The relevant quadratic Diophantine equation can be derived by considering rational values of 
$\nu_{\rm P}$ and $\Phi/\Phi_0$.  } by these considerations, but is dependent on the unique microscopic physics of each particular system. 
The gap trajectories observed in graphene on hBN include ordinary integer Hall gaps\cite{klitzing_new_1980,dean_boron_2010} with $s=0$ and $m=1$, moir{\' e} pattern band gaps with $\sigma = 0$ and $N=1$,\cite{wong_local_2015} fractional Hall gaps\cite{tsui_two-dimensional_1982,du_fractional_2009,dean_multicomponent_2011} with $s=0$ and $m > 1$, and Hofstadter gaps\cite{hunt_massive_2013,dean_hofstadter/s_2013,ponomarenko_cloning_2013,yu_hierarchy_2014,lee_ballistic_2016} with $m=N=1$ and $s \ne 0$.  
In recent work\cite{wang_evidence_2015} fractional Hofstadter states with both $s$ and $\sigma$ non-zero and $N > 1$ have been observed for the first time.  The most prominent states occur for $\sigma = 1$ and $s=1/3$ and $\sigma = 2$ and $s=-1/3$.  
The main goal of this paper is to explain this observation.

\noindent
{\em Competition between Fluid and Crystal States}--- 
Electrons in a partially filled Landau level can form Wigner crystal 
states in which translational symmetry is broken and electrons avoid each other 
by localizing collectively on lattice sites.  In the $n=0$ Landau level of a 
parabolic band two-dimensional electron system many different\cite{macdonald_broken_1985} electron crystal states can form.  The lowest energy Wigner crystal 
states at Landau level filling factors $\nu < 1/2$ are triangular lattice 
electron crystals with one electron per unit cell, whereas at Landau level 
filling factors $\nu > 1/2$ they are triangular lattice hole crystals with one 
electron missing from the full Landau level per unit cell.  
The electron crystal state has been established as the many-electron ground state for small $\nu$, and the hole crystal for $\nu$ close to $1$. 
However the ground states over the important filling factor range between $\sim 0.25$ and $\sim 0.75$, where correlation energy scales and gap sizes are larger, are fractional quantum Hall fluids characterized by quasiparticle's fractional charges and exotic statistics.\cite{jain_composite_2007} 
We explain below why this competition between crystal and fluid states 
is altered in graphene on hBN.  

\begin{figure}
\includegraphics[width=\columnwidth]{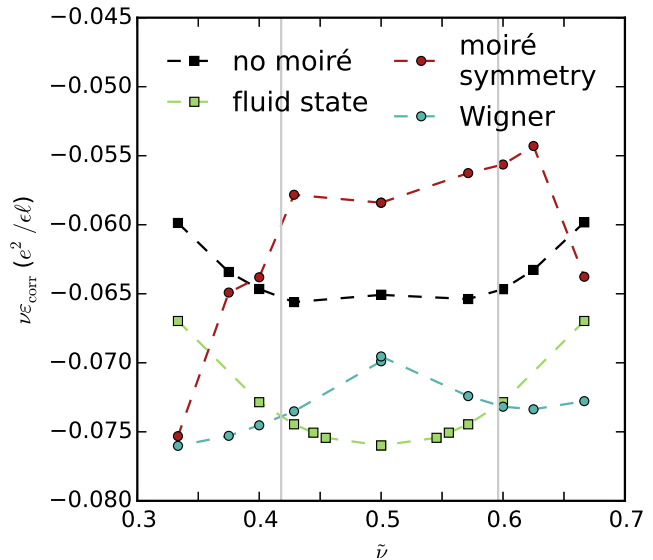}
\caption{(Color online) Correlation energy (see text) per flux quantum in $e^2/\epsilon \ell$ units {\it vs.} fractional 
level filling factor $\tilde{\nu}$ for candidate $N=0$ Landau level many-body ground states. 
($\epsilon$ is the effective dielectric constant, $\ell$ is the magnetic length and $2 \pi \ell^2 B = \Phi_0$.) 
When no moir{\' e} potential is present, the correlation energies of the Jain sequence of fluid states (green squares) lie below   
the correlation energies of the lowest energy crystal states, triangular electron crystals for $\tilde{\nu} < 1/2$ and triangular hole crystals for $\tilde{\nu}>1/2$ (black squares).  The crystal state energies 
are substantially lowered in the presence of the moir{\' e} potential (light blue circles) and fall below those of the 
fluid states over a broad filling factor range.  The dark red circles show the Hartree-Fock ground state energy in the presence of the moir{\' e} potential when the moir{\' e} pattern translational symmetry is maintained. 
These results were calculated for $\epsilon = 4$. 
Light gray vertical lines mark the cross-over filling factors, $\nu_{c,e}$ and $\nu_{c,h}$. 
}
\label{fig:energy_tot}
\end{figure}

Because crystal states spontaneously break translational symmetry, their energies have 
a first order response to a weak commensurate periodic perturbations with an energy gain 
that is optimized by choosing the most favorable position of 
the lattice relative to the periodic perturbation.  Fluid state energies, on the other hand, respond only at second order.  It follows that 
any periodic potential will favor commensurate crystal states over fluid states.  The most favorable case for inducing a transition from a fluid state to a crystal state due to a periodic potential 
is when the periodic potential has a low order commensurability with either a triangular electron crystal in a partially filled Landau 
level with $\nu < 1/2$ or a triangular hole crystal in a Landau level with $\nu > 1/2$.  Since the moir{\' e} 
pattern formed by graphene on hBN also has triangular lattice periodicity, the most 
favorable low order commensurability is one with a triangular electron crystal with one electron
located in every third moir{\' e} unit cell ($s=1/3$) or a triangular hole crystal with one hole 
located in every third moir{\' e} unit cell ($s=-1/3$). 

To understand more deeply where these broken translational symmetry states are most 
likely to appear we must closely examine the Landau levels of an isolated graphene sheet and 
their interactions with the hBN substrate.  Graphene $\pi$-band electrons are described 
at low energies by continuum models that contain both orbital and honeycomb-sublattice degrees of freedom.
For the substrate interaction contribution to the Hamiltonian, we use the spatially periodic and sublattice-dependent continuum-model derived by Jung {\it et al.}\cite{jung_origin_2015} on the basis of {\it ab initio} electronic structure calculations,\cite{jung_ab_2014} 
which is accurate when the period of the moir{\'e} pattern is larger than the
lattice constant of graphene,\cite{bistritzer_moire_2011} a condition that is safely satisfied when the graphene and substrate lattices are aligned.  The four-fold degenerate $N=0$ graphene Landau level, occupied over the Landau level filling range $-2 < \nu < 2$, consists\cite{mcclure_diamagnetism_1956,toke_fractional_2006,apalkov_fractional_2006,castro_neto_electronic_2009}  
of $n=0$ orbitals in one of two spin states and localized on opposite graphene sublattices for states in opposite valleys.  
We concentrate initially on the filling factor range $|\nu|  > 1$ over which the fractional quantum Hall physics is similar to that in a parabolic band system,
\footnote{Over the filling factor range $-1 < \nu < 1$ novel fractional quantum Hall fluids with additional stability can be established in the graphene $N=0$ Landau level by taking advantage of both spin and valley degrees of freedom.  See Inti Sodemann and Allan H. MacDonald,  Phys. Rev. Lett. {\bf 112}, 126804 (2014) and work cited therein.} 
and assume that the external magnetic field is strong enough to force maximal spin-polarization and suppress Landau level mixing by the substrate interaction Hamiltonian.

\begin{figure}
\includegraphics[width=\columnwidth]{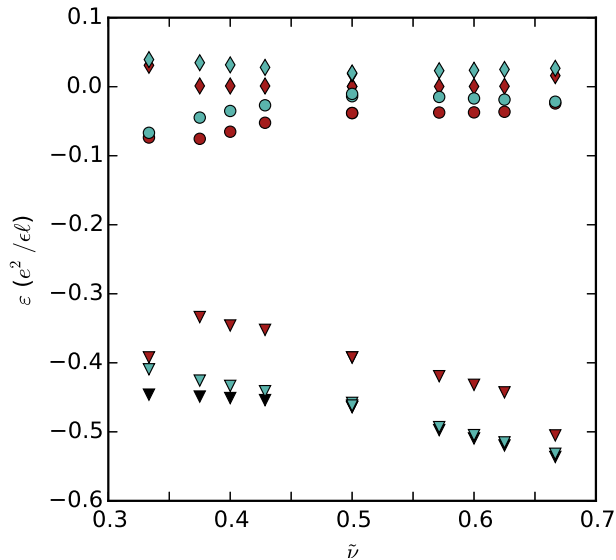}
\caption{(Color online) Contributions to the energy per electron (light blue symbols) in $e^2/\epsilon \ell$ units for electron ($\tilde{\nu} = \Phi_0/3\Phi$, $\nu < 1/2$)  
and hole ($\tilde{\nu} = 1 - \Phi_0/3\Phi$, $\nu > 1/2$) Wigner crystals.  
The circles are single-particle energies due to electron interactions with the moir{\' e} potential while the diamonds are hartree energies and triangles are exchange energies.  
The dark red symbols report results for the case in which the moir{\' e} potential translational symmetry is unbroken, while the black triangles show the exchange energy for triangular electron and hole crystalline states with no moir{\'e} potential. 
}\label{fig:energy}
\end{figure}

\begin{figure*}
\includegraphics[width=\textwidth]{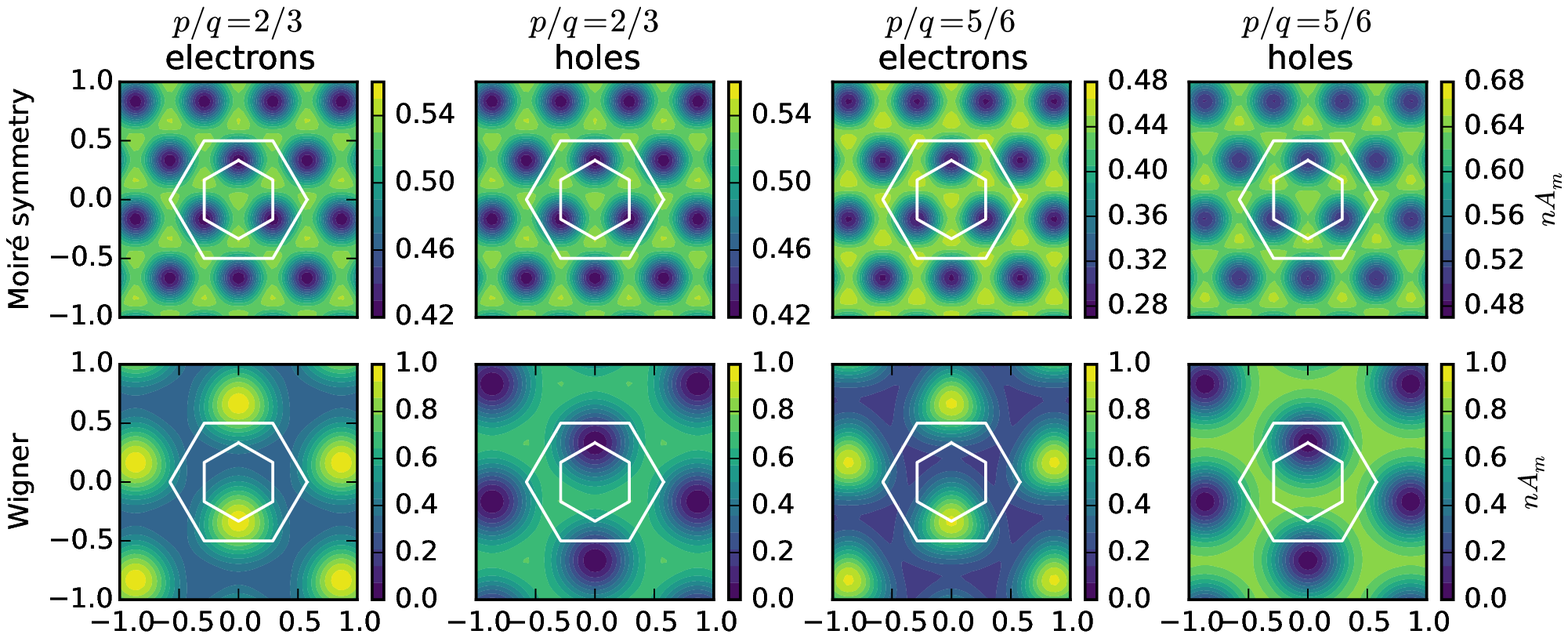}
\caption{(color online) Electron density distributions in the presence of a moir{\' e} potential predicted by single-flavor, $n=0$ Landau-level, self-consistent Hartree-Fock calculations for $\Phi/\Phi_{0}=2/3$ (two left columns) and $\Phi/\Phi_0 = 5/6$ (two right columns) at filling factors corresponding to one electron per three moir{\' e} pattern unit cells ($\nu = \Phi_0/3\Phi$ - first and third columns) and one hole per three moir{\' e} pattern unit cells ($\nu = 1 - \Phi_0/3\Phi$ - second and fourth columns).
The top row shows results obtained when the lattice symmetry defined by the moir{\' e} pattern is not broken while the lower row shows the unrestricted results obtained when the unit cell area is allowed to expand to contain three moir{\' e} pattern unit cells.
The smaller hexagons show the Wigner-Seitz cells of the moir{\' e} potentials whereas the larger hexagons show the Wigner-Seitz cells of the broken-symmetry electron and hole crystal states.
}\label{fig:density}
\end{figure*}

Because of the sublattice-localization property of the $N=0$ orbitals\cite{castro_neto_electronic_2009}, only sublattice-diagonal terms in the interaction Hamiltonian are relevant for $-2 < \nu < 2$.  
These produce an external potential that is different on different sub-lattices and hence for different valleys.  
Since the energy gained by preferentially occupying low-energy states in a broadened Landau level is larger for greater broadening, the moir{\' e} pattern on its own favors partial occupation of the valley with the stronger substrate-interaction potential.  
Additionally, tendencies toward valley and spin polarization in the fractional quantum Hall regime are usually reinforced by interactions.\cite{zhang_ground_1984,nomura_quantum_2006,apalkov_fractional_2006,zhang_landau-level_2006}
We therefore assume that for $|\nu| > 1$ only one spin-valley degree of freedom is relevant, and that the relevant valley is the one with the stronger moir{\' e} potential.  

Motivated by these considerations, we have estimated the energies of Wigner crystal states in graphene on hBN by performing self-consistent Hartree-Fock calculations in a single-component $n=0$ Landau level with an external potential chosen to be the stronger of the two sublattice potentials calculated in Ref.~[\onlinecite{jung_origin_2015}].  
The relative strength of the external potential and electron-electron interaction terms in the mean-field Hamiltonian is dependent on the effective dielectric constant.
The results reported here were obtained using $\epsilon=4$, the value that is appropriate\cite{geick_normal_1966} for graphene encapsulated by hBN.\footnote{A simple moir{\' e} pattern is formed when only one of the two adjacent layers is nearly aligned with the graphene sheet.  
The interaction between graphene $\pi$-electrons and the unaligned hBN layer is negligible.}  
The details of these calculations are outlined in Ref.~[\onlinecite{macdonald_broken_1985}]. 
Broken translational symmetry states were accessed by assuming a lattice that is less dense than but commensurate with the moir{\' e} pattern lattice and applying a seed potential at the first step of the self-consistent-field iteration process to initiate exploration of the lower symmetry solution space.   

Our results for the ground state energy as a function of the partial filling factor $\tilde{\nu}$ of the fractionally occupied Landau level are summarized in Fig.~\ref{fig:energy_tot} and compared with the energies (green squares) of the Jain sequence of fluid states\cite{jain_composite_2007} at partial filling factors  $\tilde{\nu} = n/(2n+1)$ and $\tilde{\nu} = (n+1)/(2n+1)$.
The smooth line connecting the accurately known energies of the Jain sequence states is a lower-bound for fluid state energies.  
The energies in Fig.~\ref{fig:energy_tot} are reported as correlation energies per flux quantum,
\begin{equation}
\tilde{\nu} \varepsilon_{\rm corr} \equiv \tilde{\nu} \left[\varepsilon(\tilde{\nu})-\tilde{\nu}\varepsilon(\tilde{\nu}=1)\right],
\end{equation}
a quantity that is particle-hole symmetric in the absence of external potentials.\cite{gros_conjecture_1990} 
Here $\varepsilon(\tilde{\nu})$ is the energy per electron, $\varepsilon(\tilde{\nu}=1) = \sqrt{\pi/8}\: e^{2}/\epsilon\ell$, and $\varepsilon_{\rm corr}$ is the correlation energy per electron.  
Crystal ground state energies calculated with and without moir{\'e}-pattern translational symmetry breaking are indicated by light blue and dark red circles respectively. 
For comparison, we have also plotted crystal state energies calculated in the absence of the moir{\'e} potential (black squares). 
For both fluid and crystal states, the calculations shown overestimate the ground state energy, in the fluid case because we have neglected the small second order response to the substrate interaction, and in the crystal case because of correlations neglected in the Hartree-Fock approximation.  
However, the corrections are in both cases expected to be small.  
We therefore conclude that that fluid states are lower in energy than electron Wigner crystal states for $\tilde{\nu}_{c,e}=0.418 < \tilde{\nu} < 0.5 $ and lower in energy than hole 
Wigner crystal states for $0.5 < \tilde{\nu} < \tilde{\nu}_{c,h} = 0.596$.

Since these fluid and crystalline states are formed on top of filled inert Landau levels that contribute $e^2/h$ to the Hall conductivity, their gap trajectories of fluid, electron Wigner-crystal and hole Wigner-crystal states are  
\begin{eqnarray} 
\nu_{\rm P} &=& (1 + \Sigma/m) \frac{\Phi}{\Phi_0}  \nonumber \\ 
\nu_{\rm P} &=& \frac{1}{3} + \frac{\Phi}{\Phi_0}  \nonumber \\
\nu_{\rm P} &=& - \frac{1}{3} + 2 \frac{\Phi}{\Phi_0},
\end{eqnarray} 
where $\Sigma$ and $m$ are integers that characterize the well-known variety of fractional Hall fluid states.  
It follows from Fig.~\ref{fig:energy_tot} that the second trajectory should  
emerge when $\Phi/\Phi_0 \gtrsim 0.80$, and the third trajectory when 
$\Phi/\Phi_{0} \gtrsim 0.83$.  
These findings are in close agreement with experiment, and compellingly identify the anomalous gap trajectories discovered in Ref.~[\onlinecite{wang_evidence_2015}] as evidence for Wigner crystal states that are pinned by the moir{\' e} pattern.  
We examine the properties of these states in more detail below.  

\noindent
{\em Moir{\'e}-Pattern Pinned Wigner Crystal States}--- 
The relevant sublattice-projected substrate interaction potential of graphene on boron nitride has 120$^{\circ}$ rotational symmetry with minima located on three corners of the hexagonal moir{\'e} pattern Wigner-Seitz 
cell and maxima on the other three corners.  
This property implies that the moir{\' e} potential has a unique minimum and a unique maximum in each unit cell.      
The single particle energies calculated with and without symmetry breaking are 
plotted in Fig.~\ref{fig:energy} as light blue and dark red diamonds, respectively. 
The broken-symmetry state has higher single particle energies because the system
 is not quite as effective in gaining energy from the moir{\'e} potential. 
However, this energy cost is compensated by a large reduction in the exchange energy, so that the total energy is lower. 
The exchange energies are shown in Fig.~\ref{fig:energy} as light blue and dark red triangles, respectively. 
Broken symmetry states are strongly favored when each moir{\' e} potential unit cell contains one third of an electron or one third of a hole.  
Because the moir{\'e} potential breaks particle-hole symmetry,~\cite{yankowitz_emergence_2012,ortix_graphene_2012,dasilva_transport_2015} the size of the energy gain (dark red and light blue markers) and the details of the broken symmetry state are different for electron and hole crystals, in agreement with experiment.  
Unlike the case in the absence of a periodic potential, the electron and hole Wigner crystal states do not have the same energy at $\tilde{\nu}=1/2$. 

Fig.~\ref{fig:density} shows maps of the carrier density in real space. 
The top row of this figure corresponds to two representative values of $\tilde{\nu}$ on the red line in Fig.~\ref{fig:energy_tot}, $\tilde{\nu} = 1/2$ for electrons and holes (left two plots) and $\tilde{\nu} = 2/5$ for electrons and $\tilde{\nu} = 3/5$ for holes (right two plots). 
The smaller white hexagon drawn on each of these maps shows the moir{\'e} unit cell. 
The electrons (holes) organize themselves to fill dips (peaks) in the moir{\'e} potential. 
The bottom row of Fig.~\ref{fig:density} shows the corresponding carrier density maps for the broken-symmetry states corresponding to the blue line in Fig.~\ref{fig:energy_tot},  which are also periodic but have a larger unit cell (the larger white hexagon drawn on each of the maps.) 
In this case, the carrier density maxima are close to the full Landau level density $n_{B}$. 
In comparison, the carrier density maxima in the unbroken symmetry case are $ \sim 0.5 n_{\Phi}$.

\noindent 
{\em Summary and Discussion}---
We have shown that because of the substrate interaction between graphene $\pi$-band electrons and an aligned hBN substrate, Wigner crystal states compete strongly
with the correlated fluid states that are normally associated with partially filled Landau levels.
Our theoretical results quantitatively explain recent experiments\cite{wang_evidence_2015} which discovered unexpected fractional Hofstadter butterfly trajectories in the filling factor range $1 < \nu < 2$.  We attribute the absence of similar trajectories over the corresponding negative filling factor regime $-2 \nu -1$ to particle-hole symmetry breaking by the moir{\'e} potential~\cite{yankowitz_emergence_2012,ortix_graphene_2012,dasilva_transport_2015} which leads in the absense of a magnetic field to larger optical anomalies at negative carrier densities,~\cite{dasilva_terahertz_2015} and in a magnetic field will also lead to stronger Landaul level mixing effects at negative filling factors. 
Our identification of the unexpected new states suggests that they will have as yet unstudied, but experimentally accessible, properties that further identify their character and have intrinsic interest.   
For example the Wigner crystal states that have previously been identified in nearly empty Landau level have been studied mainly by using microwave\cite{ye_correlation_2002} spectroscopy to probe their finite-frequency pinning-mode collective excitations.  In this case the pinning has been provided by random disorder potentials, and the pinning modes are broad.
For the case of graphene on hBN it should be possible for the first time to study Wigner crystal states that are pinned by a periodic perturbation and have sharp pinning-mode collective excitations.
Furthermore, there is an excellent prospect that a sliding-mode\cite{gruner_dynamics_1988} conduction mechanism will be revealed by 
careful studies of non-linear transport effects.  

\noindent
\begin{acknowledgments}
Work in Austin was supported by the Department of Energy, Office of Basic Energy Sciences under contract DE-FG02-ER45118 and by the Welch foundation under grant TBF1473. Work in Seoul was supported by the Korean NRF under grant NRF-2016R1A2B4010105. The authors would like to thank Cory Dean for useful discussions.
\end{acknowledgments}


%

\end{document}